\documentstyle[psfig]{mn}

\newif\ifAMStwofonts
                  
\def\gsim{\setbox0=\hbox{$>$}%
        \mathrel{\vtop{\baselineskip=0pt\lineskip=0pt%
        \copy0\hbox to\wd0{\hss$\scriptstyle\sim$}}}}
\def\lsim{\setbox0=\hbox{$<$}%
        \mathrel{\vtop{\baselineskip=0pt\lineskip=0pt%
        \copy0\hbox to\wd0{\hss$\scriptstyle\sim$}}}}

\def\kms{\relax \ifmmode {\,\rm km\,s}^{-1}\else \,km\,s$^{-1}$\fi}
\def\ha{\relax \ifmmode {\rm H}\alpha\else H$\alpha$\fi}
\def\hb{\relax \ifmmode {\rm H}\beta\else H$\beta$\fi}
\def\hi{\relax \ifmmode {\rm H\,{\sc i}}\else H\,{\sc i}\fi}
\def\hii{\relax \ifmmode {\rm H\,{\sc ii}}\else H\,{\sc ii}\fi}
\def\h2{\relax \ifmmode {\rm H}_2\else H$_2$\fi}
\def\lha{\relax \ifmmode L_{{\rm H}\alpha}\else $L_{{\rm H}\alpha}$\fi}
\def\shi{\relax \ifmmode \sigma_{{\rm HI}}\else $\sigma_{\rm HI}$\fi}
\def\sh2{\relax \ifmmode \sigma_{{\rm H}_2}\else $\sigma_{{\rm H}_2}$\fi}
\def\degr{\hbox{$^\circ$}}
\def\arcmin{\hbox{$^\prime$}}
\def\arcsec{\hbox{$^{\prime\prime}$}}

\def\fdg{\hbox{$.\!\!^\circ$}}
\def\fs{\hbox{$.\!\!^{\rm s}$}}
\def\farcm{\hbox{$.\mkern-4mu^\prime$}}
\def\farcs{\hbox{$.\!\!^{\prime\prime}$}}
\def\degd#1.#2{ #1\fdg#2 }                 
\def\mind#1.#2{ #1\farcm#2 }               
\def\secd#1.#2{ #1\farcs#2 }               
\def\hhh{\ifmmode {\rm ^h}              
         \else {${\rm ^h}$}
         \fi}
\def\sss{\ifmmode {\rm ^s}              
         \else {${\rm ^s}$}
         \fi}
\def\hms#1h#2m#3s{                      
                  \relax
                  \ifmmode #1^{\rm h}\,#2^{\rm m}\,#3^{\rm s}
                  \else \hbox{$#1^{\rm h}\,#2^{\rm m}\,#3^{\rm s}$}
                  \fi
                 }   
\def\dms#1d#2m#3s{                      
                  \relax
                  #1\degr\,#2\arcmin\,#3\arcsec
                 }
\def\hmsd#1h#2m#3.#4s{                  
(RA)
                      \relax
                      \ifmmode #1^{\rm h}\,#2^{\rm m}\,#3\fs#4
                      \else \hbox{$#1^{\rm h}\,#2^{\rm m}\,#3\fs#4$}
                      \fi
                     }
\def\dmsd#1d#2m#3.#4s{                  
(Dec)
                      \relax
                      #1\degr\,#2\arcmin\,#3\farcs#4
                     }
\def\mag{\relax                          
        \ifmmode ^{\rm m}
        \else $^{\rm m}$ 
        \fi
       }
\def\magd#1.#2{                          
              \relax
              \ifmmode #1^{\rm m}
                       \hskip-0.55em.\hskip0.22em#2
              \else \hbox{#1$^{\rm m}
                    \hskip-0.55em.\hskip0.22em$#2}
              \fi
             }
  
\ifoldfss
  \ifCUPmtlplainloaded \else
    \NewTextAlphabet{textbfit} {cmbxti10} {}
    \NewTextAlphabet{textbfss} {cmssbx10} {}
    \NewMathAlphabet{mathbfit} {cmbxti10} {} 
    \NewMathAlphabet{mathbfss} {cmssbx10} {} 
  \fi
  \ifAMStwofonts  
    \ifCUPmtlplainloaded \else
      \NewSymbolFont{upmath} {eurm10}
      \NewSymbolFont{AMSa} {msam10}
      \NewMathSymbol{\upi}     {0}{upmath}{19}
      \NewMathSymbol{\umu}     {0}{upmath}{16}
      \NewMathSymbol{\upartial}{0}{upmath}{40}
      \NewMathSymbol{\leqslant}{3}{AMSa}{36}
      \NewMathSymbol{\geqslant}{3}{AMSa}{3E}

    \fi
  \fi
\fi 
                                        
\ifnfssone
  \newmathalphabet{\mathit}
  \addtoversion{normal}{\mathit}{cmr}{m}{it}
  \addtoversion{bold}{\mathit}{cmr}{bx}{it}
  \newmathalphabet{\mathbfit} 
  \addtoversion{normal}{\mathbfit}{cmr}{bx}{it}
  \addtoversion{bold}{\mathbfit}{cmr}{bx}{it}
  \newmathalphabet{\mathbfss} 
  \addtoversion{normal}{\mathbfss}{cmss}{bx}{n}
  \addtoversion{bold}{\mathbfss}{cmss}{bx}{n}
  \ifAMStwofonts
    \ifCUPmtlplainloaded \else
      \UseAMStwoboldmath
      \makeatletter
      \new@mathgroup\upmath@group
      \define@mathgroup\mv@normal\upmath@group{eur}{m}{n}
      \define@mathgroup\mv@bold\upmath@group{eur}{b}{n}
      \edef\UPM{\hexnumber\upmath@group}
      \new@mathgroup\amsa@group
      \define@mathgroup\mv@normal\amsa@group{msa}{m}{n}
      \define@mathgroup\mv@bold\amsa@group{msa}{m}{n}
      \edef\AMSa{\hexnumber\amsa@group}
      \makeatother
      \mathchardef\upi="0\UPM19
      \mathchardef\umu="0\UPM16
      \mathchardef\upartial="0\UPM40
      \mathchardef\leqslant="3\AMSa36
      \mathchardef\geqslant="3\AMSa3E
    \fi
  \fi
\fi 
  
\ifnfsstwo
  \DeclareMathAlphabet{\mathbfit}{OT1}{cmr}{bx}{it}
  \SetMathAlphabet\mathbfit{bold}{OT1}{cmr}{bx}{it}
  \DeclareMathAlphabet{\mathbfss}{OT1}{cmss}{bx}{n}
  \SetMathAlphabet\mathbfss{bold}{OT1}{cmss}{bx}{n}
  \ifAMStwofonts
    \ifCUPmtlplainloaded \else 
      \DeclareSymbolFont{UPM}{U}{eur}{m}{n}
      \SetSymbolFont{UPM}{bold}{U}{eur}{b}{n}
      \SetSymbolFont{UPM}{bold}{U}{eur}{b}{n}
      \DeclareSymbolFont{AMSa}{U}{msa}{m}{n}
      \DeclareMathSymbol{\upi}{0}{UPM}{"19}
      \DeclareMathSymbol{\umu}{0}{UPM}{"16}
      \DeclareMathSymbol{\upartial}{0}{UPM}{"40}
      \DeclareMathSymbol{\leqslant}{3}{AMSa}{"36}
      \DeclareMathSymbol{\geqslant}{3}{AMSa}{"3E}
    \fi
  \fi  
\fi 
   
\ifCUPmtlplainloaded \else 
  \ifAMStwofonts \else 
    \def\upi{\pi}
    \def\umu{\mu}
    \def\upartial{\partial}
  \fi
\fi

\title{The Unified model and the Seyfert 2 infrared dichotomy}
\author[D.M. Alexander]{D.M. Alexander\thanks{email: davo@sissa.it}\\
International School for Advanced Studies, SISSA, 2-4 via Beirut, 34014
Trieste, Italy\\}

\date{Accepted;
      Received;  
      in original form}
    
\pagerange{\pageref{firstpage}--\pageref{lastpage}}
\pubyear{2000}

\begin{document}
      
\maketitle

\label{firstpage}
      
%
\begin{abstract}
%

An optical spectropolarimetric study has shown that the detectability of
polarised broad H$\alpha$ in Seyfert 2 galaxies is correlated to the IRAS
$f_{60}/f_{25}$ flux ratio where only those Seyfert 2s with ``warm" IRAS
colours show polarised broad line emission. It was suggested that those
Seyfert 2s with ``cool" IRAS colours have highly inclined tori which
obscure the broad line scattering screen. 

I present here hard X-ray observations inconsistent with this picture
showing that the derived column densities of warm and cool Seyfert 2
galaxies are statistically the same. I classify the Bright Galaxy Sample
to produce a non-Seyfert comparison. The analysis of the properties of
these galaxies with the Seyfert 2s suggest that the IRAS $f_{60}/f_{25}$
flux ratio is consistent with implying the relative strength of galactic
to Seyfert emission. I show that this new picture can account for the
absence of polarised broad H$\alpha$ in the cool Seyfert 2s.

\end{abstract}


%
\section{Introduction}
%

The unified model for Seyfert galaxies proposes that all types of Seyfert
galaxy are fundamentally the same, however, the presence of an optically
thick structure obscures the broad line region (BLR) in many systems. In
this paper it is assumed that, in the majority of Seyfert 2s, this
structure is a dusty molecular ''torus" although other galactic structures
(e.g.\ dust lanes/starbursts, see Malkan, Gorjian and Tam, 1998) can
perform the same role. In this scenario the classification of a Seyfert 1
or Seyfert 2 galaxy (Seyfert 1--broad permitted lines, Seyfert 2--narrow
permitted lines)  depends on the inclination of the torus to the line of
sight (Antonucci, 1993). Probably the most convincing evidence for this
model comes from optical spectropolarimetry. Using this technique, the
scattered emission from the BLR of many Seyfert 2s is revealed in the form
of broad lines in the polarised flux (e.g.\ Antonucci and Miller, 1985,
Young et al, 1996, Heisler, Lumsden and Bailey, 1997).

In this unified picture the high energy central source emission (optical
to X-ray continuum) is absorbed by the dust within the torus which
re-emits this energy at infrared (IR) wavelengths. Independent strong
support has been given by hard X-ray (HX, 2 to 10 keV), near-IR and mid-IR
observations (e.g.\ Turner et al, 1997, Risaliti, Maiolino and Salvati,
1999, Alonso-Herrero, Ward, Kotilainen, 1997 and Clavel et al, 2000)
showing that Seyfert 2s are generally characterised by strong absorption
whilst Seyfert 1 galaxies are relatively unabsorbed. 

Heisler, Lumsden and Bailey (1997, hereafter HLB) performed an optical
spectropolarimetric study of a well defined and statistically complete
IRAS 60$\mu$m selected Seyfert 2 sample to determine the statistical
detectability of polarised broad lines.  The objects were selected at
60$\mu$m to reduce the possibility of biasing due to torus
inclination/extinction effects and all objects were observed to the same
signal to noise to ensure similar detection thresholds. In this study a
striking relationship between the detectability of polarised broad
H$\alpha$ and the IRAS $f_{60}/f_{25}$ flux ratio was found where only
those galaxies with warm IRAS colours ($f_{60}/f_{25}<$4.0) showed a
hidden broad line region (HBLR). Both Seyfert 2 galaxy types were found to
be well matched in terms of redshift, overall polarisation and detection
rate of compact nuclear radio emission. Therefore, without any apparent
contradictory evidence, HLB suggested that the IRAS $f_{60}/f_{25}$ ratio
provides a measure of the inclination of the torus to the line of sight:
in a cool Seyfert 2 the torus is so highly inclined that even the broad
line scattering screen is obscured. I present here HX evidence that
suggests this picture is incorrect and provide a new view that is
consistent with other observations.

%
\section{Testing the inclination picture}
%

The picture presented by HLB appears reasonable. Assuming that the Seyfert
torus is optically thick at mid-IR wavelengths (e.g. Pier and Krolik,
1993, Granato and Danese, 1994, Efstathiou and Rowan-Robinson, 1995) the
mid-IR to far-IR flux ratio should vary depending upon the inclination of
the torus to the line of sight. A simple prediction of this picture is
that Seyfert 1 galaxies should show warmer colours than Seyfert 2
galaxies. The mean IRAS $f_{60}/f_{25}$ flux ratios of Seyfert galaxies
from the IRAS 60$\mu$m selected Bright Galaxy Sample (BGS, Soifer et al,
1989), as classified by Kim et al (1995) and using data from the
literature (see section 3) are 7.1$\pm$2.9 and 7.6$\pm$3.2 for Seyfert 1s
and 2s respectively. The Seyfert 1s do not statistically show warmer
colours in this sample. However, it could be argued that these ratios are
biased by differences in the star formation between Seyfert 1s and 2s or
optical depth effects in the Seyfert 2 nuclei, biasing the ratio towards
warm objects. In any case the most direct test of the inclination picture
is made with HX observations.

\subsection{The X-ray picture}

One of the key supports of the unified model comes from HX observations
where the nuclear extinction is directly determined from the observed
spectral slope. Seyfert 1 galaxies are characterised by little or no
absorption 20$<$log($N_H$)$<$21 cm$^{-2}$ whilst Seyfert 2s have
significant, sometimes extreme, absorption 22$<$log($N_H$)$<$25 cm$^{-2}$
(e.g.\ Turner et al, 1997 and Risaliti, Maiolino and Salvati, 1999). 
Although the HX properties of Seyferts are too poorly known to allow a
detailed measure of the torus inclination, as the cool Seyfert 2s are more
highly inclined than the warm Seyfert 2s in the HLB interpretation, they
should statistically show higher column densities. To date 13 of the
galaxies in the HLB sample have been observed with either BeppoSAX or
ASCA. The other 3 objects have been observed by Einstien or in the HEAO1/A
survey. In the case of the HEAO1/A objects only upper limits could be
placed. For these two galaxies (NGC34 and NGC1143) I have used the upper
limits and unextincted [OIII]$\lambda$5007 emission line fluxes to predict
their nuclear extinction using the diagnostic diagram of Bassani et al
(1999).  The distribution of HX derived column densities are shown in
figure 1 and presented in table 1.

%
%

\begin{table*}
\caption{\em The HLB Seyfert 2 sample properties.}
\label{tab:table}
\begin{center}
\leavevmode
\footnotesize
\begin{tabular}{lclcccr}
\hline \\[-5pt]
Galaxy & z & $f_{H\alpha}$ & $f_{60}$/$f_{25}$ & 
$H_{\alpha}$/$H_{\beta}$ & HBLR? & $N_H$ \\[+5pt]
\hline \\[-5pt]
NGC0034      & 0.01978 & -14.9$_p$ & 7.01 & 25.0 & No & $\gsim$23.0$_d$\\
NGC1068      & 0.00379 & -12.7 & 2.07 & 6.2 & Yes & $>$24.0$_a$\\
NGC1143      & 0.02822 & -15.0$_p$ & 8.37 & 11.0 & No & $\gsim$22.0$_d$\\
I05189-2524  & 0.04256 & -14.5 & 3.97 & 5.8 & Yes & 22.7$_a$\\
NGC4388      & 0.00842 & -14.8 & 2.96 & 5.5 & Yes & 23.6$_a$\\
IC3639       & 0.01096 & -14.1$_p$ & 3.32 & 4.6 & Yes & $>$24.0$_c$\\
I13197-1627  & 0.01718 & -14.0 & 2.06 & 4.8 & Yes & 23.9$_a$\\
NGC5135      & 0.01372 & -14.4$_p$ & 7.04 & 7.1 & No & $>$24.0$_a$\\
I19254-7245  & 0.06171 & -15.1$_p$ & 4.42 & 10.0 & No & 23.3$_b$\\ 
IC5063       & 0.01135 & -13.3 & 1.36 & 5.9 & Yes & 23.4$_a$\\
NGC7130      & 0.01615 & -15.4$_p$ & 7.78 & 8.7 & No & $>$24.0$_c$\\
NGC7172      & 0.00859 & -16.4$_p$ & 7.50 & 6.8 & No & 22.9$_a$\\
NGC7496      & 0.00550 & -15.4$_p$ & 5.28 & 5.6 & No & 22.7$_e$\\
NGC7582      & 0.00525 & -14.1$_p$ & 7.63 & 8.9 & No & 23.1$_a$\\
NGC7590      & 0.00532 & -16.5$_p$ & 8.32 & 17.8 & No & $<$20.9$_a$\\
NGC7674      & 0.02901 & -14.9 & 2.95 & 4.4 & Yes & $>$24.0$_a$\\
\hline \\
\end{tabular}
\end{center}
{\em Notes. The H$\alpha$ emission line flux is in
log(ergs$^{-1}$cm$^{-2}$) and refers to the observed polarised broad
H$\alpha$ flux (p=predicted flux); the HBLR refers to whether polarised
broad H$\alpha$ emission has been detected; $N_H$ in log(cm$^{-2})$.
References. (a) Bassani et al, 1999, (b) Pappa, Georgantopoulos and
Stewart, 1999, (c) Risalti, Maiolino and Salvati, 1999, (d) Polletta et
al, 1996, (e) Kruper, Urry and Canziares, 1990}
\end{table*}


%
%

\begin{figure}
\begin{center}
\leavevmode
\centerline{\psfig{figure=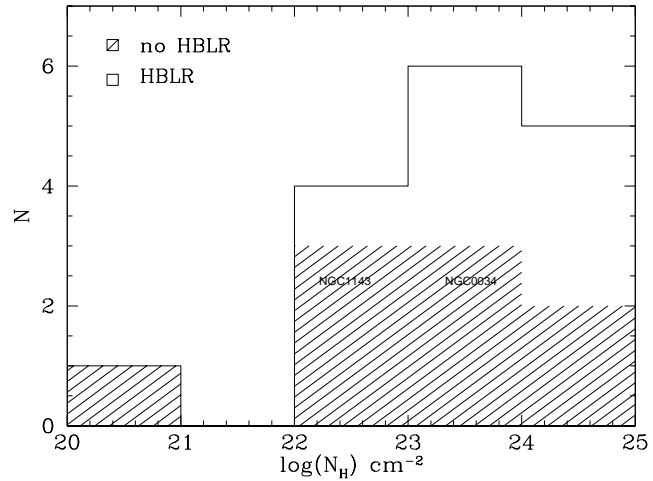,width=9cm,angle=-90}}
\vspace{0.5cm}
\end{center}
\caption{\em The distribution of hard X-ray derived hydrogen column   
densities for the Seyfert 2 galaxies in the HLB sample. The uncertain
column densities of NGC0034 and NGC1143 are highlighted.}
\label{fig:sample1}
\end{figure}


The derived column densities show that an optically thick structure exists
in both the warm and cool Seyfert 2 galaxy types, with the notable
exception of NGC7590 which may be a Seyfert 1 with galactic dust obscuring
the BLR, however, signficantly there is very little difference in the
distribution of column densities. The overall distribution is similar to
that found for the [OIII]$\lambda$5007 selected Seyfert sample of
Risaliti, Maiolino and Salvati (1999) suggesting that the far-IR selects
Seyferts in a reasonably unbiased manner: $\sim$35\% of the objects are
Compton thick (i.e. log($N_H)>24$ cm$^{-2}$), the mean log($N_H$) for the
whole sample is 23.2$\pm$0.9 cm$^{-2}$ and the mean for the warm and cool
Seyfert 2s are 23.7$\pm$0.5 cm$^{-2}$ and 22.9$\pm$1.0 cm$^{-2}$
(23.1$\pm$0.7 cm$^{-2}$ if NGC7590 is excluded) respectively. The
hypothesis of HLB could still be retained if the cool Seyfert 2s are
Compton thick and the determined column densities refer to the extinction
suffered by the scattered emission, however, the mean log([OIII]/HX) of
0.3$\pm$1.0 and -0.2$\pm$1.4 for the warm and cool Seyfert 2s respectively
suggest that this is not the case (see Bassani et al, 1999). Therefore the
most direct conclusion to make from this HX analysis is that the IRAS
$f_{60}/f_{25}$ flux ratio does not indicate the inclination angle of the
torus in Seyfert 2s.

%
\section{The Seyfert 2 infrared dichotomy}
%

If the IRAS $f_{60}/f_{25}$ colour ratio is not an indicator of the
inclination of the dusty torus, what does this colour ratio imply? A
natural starting point is to compare the HLB Seyfert properties to those
of non-Seyfert galaxies. A good comparison is the BGS sample which is
selected at the same wavelength as the HLB sample (60$\mu$m) and has a
very similar flux limit. The BGS sample is partially classified by Kim et
al (1995) using the optical emission line ratio technique (e.g. Baldwin,
Phillips and Terlevich, 1981). To increase the number of classified
objects I have taken these observations and other optical spectroscopic
observations from the literature, classifying 77\% of the BGS sample, see
table 2. These galaxies have been classified using the emission line
diagnostics of Veilleux and Osterbrock (1987) and the mode classification
for each galaxy is adopted. For brevity only the
[NII]$\lambda$6583/H$\alpha$ vs [OIII]$\lambda$5007/H$\beta$ diagram is
shown here, see figure 2.

%
%

\begin{table}
\caption{\em Classification of the BGS sample.}
\label{tab:table}
\begin{center}
\leavevmode
\footnotesize
\begin{tabular}{cccc}
\hline \\[-5pt]
HII & LINERs & AGN & No emission lines\\[+5pt]
\hline \\[-5pt]
62\% & 25\% & 12\% & 1\% \\
\hline \\
\end{tabular}
\end{center}
\end{table}


%
%
 
\begin{figure}
\begin{center}
\leavevmode
\centerline{\psfig{figure=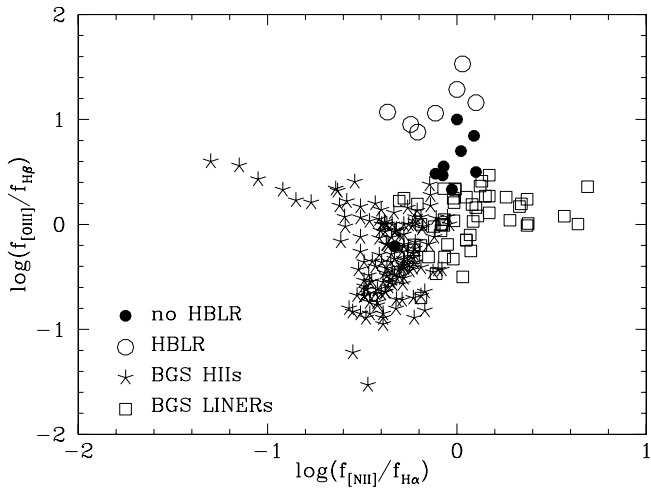,width=9cm,angle=-90}}
\vspace{0.5cm}
\end{center}
\caption{\em The optical emission line ratios for LINERs and HII galaxies
from BGS and Seyfert 2 galaxies from HLB.}
\label{fig:sample1}
\end{figure}


\subsection{The IRAS picture}

A logical first step is to compare the IRAS colours, see figure 3. The IR
warm region is clearly dominated by Seyfert 2s although the cool region
also contains HII and LINER galaxies with a wide range of IRAS colours. 
The cool Seyfert 2s cannot be distinguished from the HII and LINER
galaxies in terms of their IRAS emission (note there are no Seyfert 2s
with log($f_{60}/f_{25})>$0.9 due to the HLB selection criteria). The
distribution of HX derived nuclear column densities suggest an optically
thick structure (i.e.\ the torus) exists in both Seyfert 2 galaxy types
which, according to the unified model, should emit thermally at IR
wavelengths. Therefore, although the HX emission shows that a Seyfert
nucleus is present in the cool Seyfert 2s, the IR emission from the torus
must be dominated by galactic emission in the large IRAS apertures (as
previously suggested by Alexander et al, 1999).  Additional evidence for
this picture is found in the distribution of optical emission line ratios
where the cool Seyfert 2s have, on average, weaker [OIII]/H$\beta$
emission, see figure 2. Assuming that both the warm and cool Seyfert 2s
have the same basic Seyfert nucleus and galactic emission, the lower mean
emission line ratio in the cool Seyfert 2s implies a larger ratio of
galactic to Seyfert activity. Indeed in one galaxy (NGC7496) the observed
emission line ratio is consistent with that of an HII galaxy even though
it clearly has HX emission and therefore a Seyfert nucleus.

%
%

\begin{figure}
\begin{center}
\leavevmode
\centerline{\psfig{figure=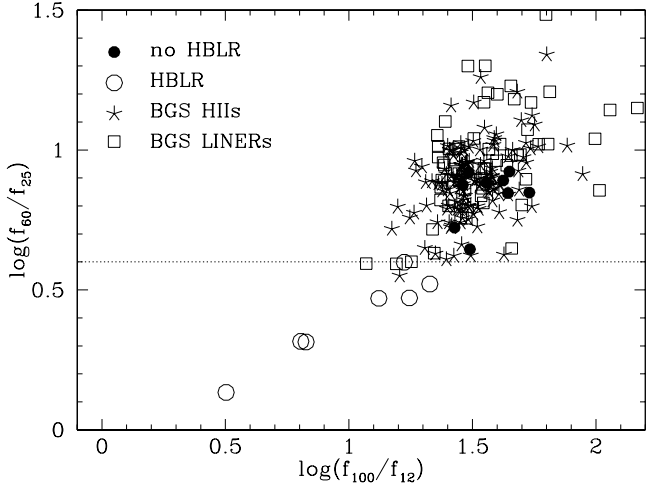,width=9cm,angle=-90}}
\vspace{0.5cm}
\end{center}
\caption{\em The IRAS colour distribution for LINERs and HII galaxies from
BGS and Seyfert 2 galaxies from HLB. The warm/cool divide is indicated by
the dotted line.}
\label{fig:sample1}
\end{figure}


\subsection{The PAH picture}

HII galaxies produce mid-IR PAH emission (see Puget and Leger, 1989) 
whilst pure Seyfert nuclei do not. Therefore in the above scenario the PAH
equivalent width should vary based on the IRAS $f_{60}/f_{25}$ colour
ratio.  Indeed Lutz et al (1998) found that this occurs in the extremely
luminous object class of ULIRGs although this test has not been performed
for Seyfert 2 galaxies. Unfortunately suitable mid-IR observations are not
available for the complete HLB sample although Seyfert 2 observations from
Clavel et al (2000) and HII galaxy observations from Rigopoulou et al
(1999) provide the general picture, see figure 4.

%
%

\begin{figure}
\begin{center}
\leavevmode
\centerline{\psfig{figure=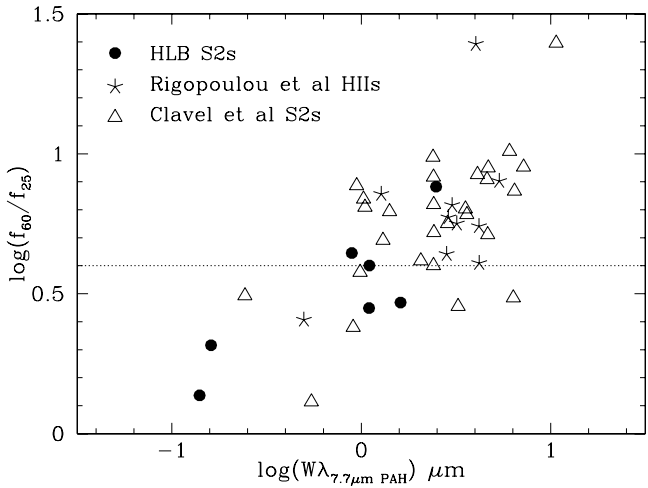,width=9cm,angle=-90}}
\vspace{0.5cm}
\end{center}
\caption{\em The distribution of observed 7.7$\mu$m PAH equivalent widths.
The warm/cool divide is indicated by the dotted line.}
\label{fig:sample1}
\end{figure}


The distribution of PAH emission is reasonably well segregated with the
cool Seyfert 2s occupying a similar region to the HII galaxies although
there is not a tight correlation. This may be because the galactic
emission can be a combination of both galactic cirrus emission, which show
strong PAH features, and more intense star formation emission, which show
weak PAH features (e.g.\ Laurent et al, 1999).

%
\section{A new explanation of the spectropolarimetric results?}
%

The cool Seyfert 2s have IRAS colours, optical emission line ratios and
PAH emission suggestive of a weak Seyfert nucleus and a dominant galactic
component. Can the spectropolarimetric results be explained in the same
way?

\subsection{HLBR sensitivity}

As the [OIII]$\lambda$5007 emission line is a good indicator of the
strength of the Seyfert nucleus (e.g. Alonoso-Herrero et al, 1997, Bassani
et al, 1999) it should be possible to estimate the strength of polarised
broad H$\alpha$ in the cool Seyfert 2s. Indeed the [OIII]$\lambda$5007
emission line flux, corrected for galactic extinction and then
re-extincted to the wavelength of H$\alpha$ to simulate the probable
extinction suffered by the H$\alpha$ emission, shows a good correlation
with polarised broad H$\alpha$ flux, see figure 5 and table 1. In general
the predicted fluxes of the cool Seyfert 2s are a factor $\sim$3 lower
than the warm Seyfert 2s and in many cases fall below that easily
obtainable on a 4 metre telescope. In two cases the predicted fluxes are
similar to those found for warm Seyfert 2s and in one of these objects
(NGC7582) broad H$\alpha$ has been directly observed (Aretxaga et al,
1999).

%
%

\begin{figure}
\begin{center}
\leavevmode
\centerline{\psfig{figure=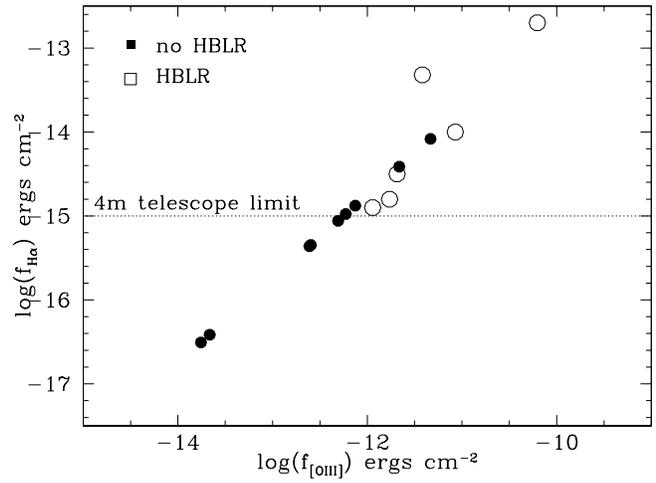,width=9cm,angle=-90}}
\vspace{0.5cm}
\end{center}
\caption{\em Correlation between the polarised broad H$\alpha$ flux and
the [OIII]$\lambda$5007 flux adjusted as if observed at the wavelength
of H$\alpha$ (see text) for the Seyfert 2s in the HLB sample.}
\end{figure}


However, in determining the detectability of polarised broad H$\alpha$
there is another factor to consider as the galaxies were observed to the
same signal to noise and therefore the detectability of a HBLR also
depends on the optical brightness of the galaxy. The mean optical (V band)
magnitudes are 12.1$\pm$2.4 and 12.4$\pm$1.4 for the warm and cool Seyfert
2s respectively, showing that there is no statistical difference in their
overall optical brightnesses. Therefore as the polarised broad H$\alpha$
fluxes are lower in the cool Seyfert 2s, whilst the overall fluxes are
not, the sensitivity to polarised broad H$\alpha$ should be worse by a
factor of $\sim$3 in the cool Seyfert 2s.

These two factors can account for the lack of detected HBLRs in the cool
Seyfert 2s and agree with the overall hypothesis that the differences
between the Seyfert 2 galaxy types are that the cool Seyfert 2s have a
weak Seyfert nucleus and a dominant galactic component.

\subsection{The measured polarisation}

The weaker Seyfert emission in the cool Seyfert 2s should lead to a lower
observed polarisation, however, HLB claim no statistical difference
between the Seyfert 2 galaxy types. This apparent inconsistency can be
reconciled as there is another component to the observed polarisation:
galactic polarisation. Galactic polarisation is proportional to the
measured galactic extinction (Serkowski, Mathewson and Ford, 1975 and
Scarrott, 1996) and as the cool Seyfert 2s have a higher galactic
extinction (i.e. as measured by the H$\alpha$/H$\beta$ emission line ratio
(Ward et al, 1987), see HLB and table 1) they should also have a larger
galactic polarisation component. 

An obvious question to raise is why do the cool Seyfert 2s have a higher
galactic extinction? A natural explanation for this would be that the
galactic disc is more highly inclined in the cool objects, however the
mean inclinations (measured from DSS images) of 43$\pm$30 and 42$\pm$25
degrees for the warm and cool Seyfert 2 galaxies respectively show that
this is not the case. Therefore there must be another component
contributing to the higher extinction in the cool Seyfert 2s. Analysis of
DSS images of the HLB sample suggest that cool Seyfert 2s are more likely
to have nuclear dust lanes. Dust lanes are known to be very efficient at
causing galactic polarisation (see Centaurus A, Hough et al, 1987) and
would also account for the higher galactic extinction in the cool Seyfert
2s. High resolution images of the HLB sample are required to test this
scenario.

%
\section{Conclusions}
%

HLB showed a dichotomy in the detectability of polarised broad H$\alpha$
in Seyfert 2 galaxies where only those objects with warm IRAS colours
showed polarised broad line emission. They suggested that those Seyfert 2s
with cool IRAS colours had highly inclined tori which obscured the broad
line scattering screen. I have presented evidence against this picture and
suggested that the IR dichotomy is due to the relative strength of
galactic to Seyfert emission. The main results of this study are: 

(i) the hard X-ray derived column densities of the warm and cool Seyfert
2s are statistically the same and therefore the IRAS $f_{60}/f_{25}$ flux
ratio cannot imply the inclination of the torus in the standard unified
model.

(ii) the cool Seyfert 2s have IRAS colours and PAH emission similar to HII
galaxies and optical nuclear emission line ratios consistent with a
mixture of Seyfert and galactic activity. As the hard X-ray observations
suggest that both Seyfert 2 galaxy types have similar nuclear properties,
the differences in the IRAS $f_{60}/f_{25}$ flux ratios are most probably
due to a weaker Seyfert component and a more dominant galactic component
in the cool Seyfert 2s.

(iii) the difference in the detectability of polarised broad H$\alpha$ is
most probably due to a larger component of optical starlight in the cool
Seyfert 2s. The reason for the lack of difference in observed polarisation
between both Seyfert 2 types is probably due to a larger galactic
polarisation component in the cool galaxies caused by their larger
galactic extinction. 

\section*{Acknowledgements} I thank the TMR network (FMRX-CT96-0068) for
support through a postdoctoral grant. This research has made use of the
NASA/IPAC Extragalactic Database (NED), which is operated by the Jet
Propulsion Laboratory, California Instutite of Technolody, under contract
with NASA. I am extremely grateful to the many people who have commented
on earlier drafts of this paper particularly Prof. Matt Malkan, the
referee, and Dr. Stuart Young. I am in debt to the late Dr. Charlene
Heisler for interesting and enthusiastic discussions whilst working on
this topic.

%


\bsp

\label{lastpage}


\end{document}